\def\eps0{\ensuremath{\epsilon _0}}
\def\2pieps0{\ensuremath{2\pi \epsilon _0}}
\def\4pieps0{\ensuremath{4\pi \epsilon _0}}

\documentclass[11pt]{article}
\usepackage{epsfig}
\usepackage{alltt}
\usepackage{amsmath}
\usepackage{listings}
\usepackage{braket}
\title{Binary Subdivision for Quantum Search}
\author{M Nordin Zakaria\\
  High Performance Computing Center,\\
  Universiti Teknologi PETRONAS,\\
  Perak, Malaysia\\
  \texttt{nordinzakaria@petronas.com.my}}
\date{}
\begin{document}
\maketitle
\bibliographystyle{plain}

\section {Introduction}

More than a decade ago, Grover \cite{grover1996} discovered a quantum algorithm for searching
unsorted list runs in $O(\sqrt{N})$.
The algorithm is important as the best that can be attained classically
with an unsorted list is in the order of $O(N)$.
Since the publication of the work, there have been various improvements made to
Grover's algorithm. Some work (\cite{brassard98},\cite{durr96},\cite{hoyer_neerbek2001},\cite{nayak1998}) extends the result to other search-based algorithms, while others (\cite{Cerf1998},\cite{grover05},\cite{grover08},\cite{tucci2010}) focusses on the nature of the
unitary transformation involved in the search algorithm.  The latter research direction is to be expected as in the words of Grover \cite{grover97}, the core of the algorithm is the "design of the unitary evolution of the system".

Though Grover's algorithm clearly outperforms its classical counterpart,
it is still not 'fast' enough when applied to NP-complete problems.
A straightforward application to the Travelling Salesman Problem, for example, following the approach in \cite{durr96},
results in an $O(\sqrt{n!})$ performance, where $n$ is the number of cities.
The result is still not computationally tractable for large values of $n$.
This is surprising, as due to the inherent parallelism in quantum computation,
one would expect a better result.
Hence, in this paper, instead on focussing on the design of the unitary search evolution, we focus instead on the possibility of embedding a quantum search algorithm within a classical binary search framework.
The result appears promising: taking full advantage of quantum parallelism, we show that it may actually be possible to search an unstructured list in $O(lg(N))$, provided we are willing to restart the quantum search multiple times with a different sequence of qubits and perform a series of measurements at the end of each.

The gist of the idea in our research approach comes from a
classical algorithm  within Computer Graphics and the
Computational Geometry: ray tracing.
In ray tracing, a ray is casted into a scene comprising of a number of objects.
A brute force approach would test the ray against each and every object in the scene,
a process that scales linearly with the number and the complexity of the objects.
A smarter approach requires that the scene be partitioned into regions,
and the bounding volume of each region be approximated. For each region,
the ray would then be tested first against the bounding volume approximation.
If the test is positive, then only would the ray be tested against each object
within the region. Can a similar scheme be invented for Grover's algorithm?
We would like to partition the list of items being searched into sublists.
For each sublist, before we even perform repetitive amplification,
we first check as to whether or not the desired item is somewhere within it.
If the item is not there, we move on to the next sublist, and repeat the same query. We elaborate on this idea in the next section.

\section {Binary Subdivision Algorithm}

Our algorithm can be intuitively understood in classical terms as follows:
We wish to search for an item, $t$, in an unsorted list, $L$.
Suppose we are given a 'magic' function, $F$, that given $L$ as input, instantaneously provide an output that indicates whether or not $t$ is within the list. $F$ does not pinpoint the location of $t$;
it merely says whether or not $t$ is somewhere within the input list.
We can then search using the following procedure:

\textit{
Search(L)
\begin{enumerate}
\item
Divide $L$ into two sublists: $L_1$ and $L_2$.
\item
Using $F$, check whether $L_1$ contains $t$. If yes, call Search($L_1$).
If not, call Search($L_2$).
\end{enumerate}
}
Assuming that $t$ exists in $L$, the recursive procedure will lead to $t$.
We claim that $F$ can be implemented as a quantum function, and that the overall procedure can be run classically.

As in Grover's algorithm, we start with 2 registers: $n$ qubits in the first,
and $1$ qubit in the second.
Note that the $n$ qubits in the first register represents $N = 2^n$ numbers.
In Grover's algorithm, the first register is initialized to be in the state $\ket{Q} = \ket{0}^{\otimes n}$, and the Hamadard operator, $H^{\otimes n}$, applied to it. The result is a linear combination of all $2^n$ computational basis states
$\ket{S} = \frac{1}{\sqrt{N}}\sum_{i=0}^{N-1}\ket{i}$. The second register is initialized to $\ket{1}$, and after a Hamadard transformation, comes to be in the state $\ket{-}$. Its sole purpose is to serve as the second qubit for the $I_t$ oracle operator.
Repeated application of Grover operator, $G = -H I_0 H I_t$, is performed to increase the probability of obtaining the desired state, $\ket{t}$, when a measurement is made upon the first register. The $I_t$ in the operator inverts the phase of $\ket{t}$. The $I_0$ operator inverts $\ket{0}$. Altogether, Grover operator works by amplifying the phase of the desired item, $\ket{t}$.

Instead of performing a quantum search on the entire list of items within $\ket{S}$, we wish to search only where $\ket{t}$ might be. Our approach requires two main features:
\begin{enumerate}
\item
the ability to determine whether or not a state $\ket{S}$ contains $\ket{t}$. The function $F$ described in the first paragraph of this section has this ability.
\item
the ability to partition the list of items being searched into sublists.
\end{enumerate}

The first feature is implemented by first running the oracle $I_t$ in such a way that its second register is $\ket{1}$ for
a target item, and $\ket{0}$ otherwise. Assuming that at most there can be only one target item,
the resulting state of the second register will then be either $\ket{0}$ or
$\sqrt{1-1/N}\ket{0} + 1/\sqrt{N}\ket{1}$. The distance between the 2 possible results decreases exponentially with increasing value of $N$. We show however how a series of non-unitary measurement operators with \textit{non-negligible} probabilities of success can be used to differentiate between the two possible states.

As for the second feature, we note that the items in Grover's search do not actually exist as in a classical, conventional list. Instead the list is actually a single state, $\ket{S}$, that encodes a superposition of possible item values.
We can however perform a partitioning by fixing qubits within $\ket{Q}$ and appropriately modifying the form of the Hamadard transform applied to it. As a specific example, consider the leftmost qubit, that is the one with the most significant digit. Half of the items in the state $\ket{S}$ that result after a $H^{\otimes n}$ operation on $\ket{Q}$ will start with $\ket{0}$ and the other half with $\ket{1}$. But if we fix the leftmost qubit in $\ket{Q}$ to be $\ket{0}$ (or $\ket{1}$) and perform a $I\otimes H^{\otimes n-1}$ instead, then the state $\ket{S}$ will only consists of items that start with $\ket{0}$ (or $\ket{1}$).

A more elaborate, formal elaboration of the proposed algorithm is then as follows:
Let the initial state of the first register  be $\ket{Q} = \ket{q_0}\ket{q_1}...\ket{q_{n-1}}$.
Assume that we are now determining the value of the $k^{th}$ qubit within the first register, where $0 \le k < n$.
Let $b$ be a binary variable that contains the state value ($\ket{0}$ or $\ket{1}$) to be tried out for $q_k$.
Let $\ket{Q_0}$ be the sublist of $\ket{Q}$ for which the qubit values have already been determined.
The general form of $\ket{Q}$ is then as follows: $(\ket{Q_0}\ket{q_k}\ket{q_{k+1}}...\ket{q_{n-1}})$.
Define as usual the function, $f:{\ket{0},...,\ket{N-1}}\rightarrow {0,1} $, that recognizes the solution:
\begin{equation}
    f(\ket{i}) =
    \begin{cases}
        1 & \text{if } \ket{i} == \ket{t} \\
        0 & \text{otherwise}
    \end{cases}
\end{equation}
Our algorithm then proceeds as follows:
\begin{enumerate}
\item
Initialize $\ket{Q_0}$ to be a null list ($\ket{}$), and set $k=0$, and $b=0$.
Initialize as well the second register to $\ket{0}$.
\item
Repeat the following steps:
\begin{enumerate}
\item
Prepare the initial state $\ket{Q}= \ket{Q_0}\ket{b}\ket{0}\ket{0}...\ket{0}$.
\item
Perform $I^{\otimes k}\otimes H^{\otimes (n-k)}$ on $\ket{Q}$:
\begin{equation}
I^{\otimes k}\otimes H^{\otimes (n-k)}(\ket{Q_0}\ket{b}\ket{0}\ket{0}...\ket{0}) = \ket{Q_0}\ket{b}\frac{\sum_{i=0}^{2^{n-k-1}}\ket{i}}{\sqrt{2^{n-k-1}}} = \ket{\psi}
\end{equation}
\item
Perform $I_t$ on $\ket{\psi}\ket{0}$:
\begin{align}
I_t{\ket{\psi}\ket{0}} & = I_t\left(\ket{Q_0}\ket{b}\frac{\sum_{i=0}^{2^{n-k-1}}\ket{i}}{\sqrt{2^{n-k-1}}}\ket{0}\right)\\
                        &= \ket{Q_0}\ket{b}\frac{\sum_{i=0}^{2^{n-k-1}}\ket{i}\ket{0 \oplus f(\ket{Q_0}\ket{b}\ket{i}) }}{\sqrt{2^{n-k-1}}}
\end{align}
\item
Consider the 2 possible results when the second register is measured:
\begin{itemize}
\item
If the target item is not within $\ket{\psi}$, then the result of the observation will be $\ket{y} = \ket{0}$.
Otherwise, it will be
\begin{equation}
\ket{x} = \sqrt{1-\frac{1}{N}}\ket{0} + \frac{1}{\sqrt{N}}\ket{1}
\end{equation}
where $N = 2^n$.
Assume for now that it is possible to differentiate between $\ket{x}$ and $\ket{y}$.
Hence, we can proceed as follows:
\begin{itemize}
\item
Suppose the target item is not within $\ket{\psi}$.
Then if $b==0$, we set $b = 1$ and repeat step 2.
\item
If the target item is within $\ket{\psi}$,
we then append $\ket{q_k}$ to $\ket{Q_0}$ - its state value considered fixed -, and proceed on by repeating step 2 with $b=0$ and $k=k+1$.
\end{itemize}
\end{itemize}

\end{enumerate}
\end{enumerate}

The complexity of the above algorithm, in terms of the number of queries to the oracle, assuming that we perform recursive subdivision without actually doing any amplitude amplification, is obviously $lg(2^n)$.

\section{Differentiating between $\ket{x}$ and $\ket{y}$} \label{differentiate}

We need to differentiate between
\begin{equation}
\ket{x} = \sqrt{\frac{N-1}{N}}\ket{0} + \frac{1}{\sqrt{N}}\ket{1}
\end{equation}
and
\begin{equation}
\ket{y} = \ket{0}
\end{equation}

The dot product between the two however approaches 1 with increasing $N$, that is
$\sqrt{(N-1)/N} \rightarrow 1$ when $N \rightarrow \infty$.
Nevertheless, if it is possible to apply a transformation using the following matrix, we would be able to tell the two apart:
\begin{equation}
D =
  \left(
    \begin{array}{cccc}
    1     &   -\sqrt{N-1} \\
    0 	    &  \sqrt{N}    \\
    \end{array}
  \right)
\end{equation}

Note that:
\begin{equation}
D \ket{x} = \ket{1}
\end{equation}
and
\begin{equation}
D \ket{y} = \ket{0}
\end{equation}

Unfortunately, $D$ is not a unitary matrix.
It is however possible to apply a non-unitary operator as a quantum measurement operator, as proposed in the work
by Terashima and Ueda \cite{terashima2005}. We need to convert $D$ into an operator that is implementable
as a quantum operator. To do this we perform the following transformation steps:

\begin{enumerate}
\item
We first factor $D$ using Singular Value Decomposition (SVD):
\begin{equation}
D = Q V R^{T}
\end{equation}
$Q$ and $R$ are unitary matrices, while $V$ is a diagonal matrix.
$Q$ and $R$ can be implemented easily within a unitary quantum framework.
We are left with $V$, a $2\times 2$ diagonal matrix that comes with one very large and one very small diagonal value.
For example, if $n=10$, $V$ would be:
\begin{equation}
V =\left(
   \begin{array}{cccc}
  45.249308037472204  & 0 \\
                   0  &  0.707193134832027
   \end{array}
   \right)
\end{equation}

And if $n=30$, $V$ would be:
\begin{equation}
V =\left(
   \begin{array}{cccc}
   46340.95000644678  & 0   \\
                   0  & 0.70710678127
   \end{array}
   \right)
\end{equation}

\item
We now factor $V$ into its roots, such that the first diagonal element becomes a value less than 2.
The power of the root depends on the size of the first diagonal element.
If we take the $16^{th}$ root of $V$, for example, we obtain
\begin{equation}
V = V_{1/16} V_{1/16} ... V_{1/16} 
\end{equation}
where $V_{1/16}$ is applied 16 times.

In the case of $n=30$, $V_{1/16}$ would be as follows:
\begin{equation}
V_{1/16} =\left(
   \begin{array}{cccc}
  1.957144124161160   &                0 \\
                   0  & 0.978572062094820
   \end{array}
   \right)
\end{equation}

\item
We then normalize the diagonal elements of $V_{1/16}$.
Let $M_0$ be the resulting matrix.
In the case of $n=30$, $V_{1/16}$ matrix would then become
\begin{equation}
M_0 =\left(
   \begin{array}{cccc}
   0.894427190997313   &                0   \\
                   0   & 0.447213595505164
   \end{array}
   \right)
\end{equation}

\end{enumerate}

The sequence of transformation to be applied to either $\ket{x}$ or $\ket{y}$ is then in the following order:
\begin{enumerate}
\item
$R^{\dagger}$
\item
$M_0 ... M_0$ ($v$ times where $v$ indicates the power of the root of $V$ applied to obtain $V_{1/v}$)
\item
$Q$
\end{enumerate}

We prepare the matrices used in the above sequence for each of the possible values of $n$ encountered in the quantum binary search in Section 2.

Note that the measurement operator complementary to $M_0$ is computed as follows \cite{terashima2005}:
\begin{equation}
M_1 =  \sqrt{1-M_{0}^{\dagger}M_0}
\end{equation}
The effect of $M_0$ on a state $\ket{\psi}$ is computed as follows:
\begin{equation}
\ket{\psi} \rightarrow \frac{M_0 \ket{\psi}}{\sqrt{\braket{\psi | M_{0}^{\dagger}M_{0} | \psi}}}
\end{equation}

\subsection {Example}

We consider as a complete numerical example, the case when $n=20$.
Suppose the target item is within the state in the first register.
The state of the second register will then be:
\begin{equation}
x = 0.999999523162728\ket{0} + 0.000976562500000\ket{1}
\end{equation}

The matrix D that we want is:
\begin{equation}
D =\left(
   \begin{array}{cccc}
   1 & -1023.999511718634   \\
   0 &  1024.0
   \end{array}
   \right)
\end{equation}

Performing SVD, we obtain $D = QVR^{T}$ where
\begin{equation}
Q =\left(
   \begin{array}{cccc}
  -0.707106781186547 & -0.707106781186547 \\
   0.707106781186547 & -0.707106781186547
   \end{array}
   \right)
\end{equation}

\begin{equation}
V =
\left(
   \begin{array}{cccc}
   1448.154515236507 &  0 \\
                   0 &  0.707106865480
   \end{array}
   \right)
\end{equation}
and
\begin{equation}
R =
   \left(
   \begin{array}{cccc}
  -0.000488281308208 & -0.999999880790675  \\
   0.999999880790675 & -0.000488281308208
   \end{array}
   \right)
\end{equation}

Now,
\begin{equation}
V^{1/16} = 
   \left(
   \begin{array}{cccc}
   1.575980833365910  &                 0   \\
                   0  & 0.978572069378633
   \end{array}
   \right)
\end{equation}

Normalizing the matrix (along the diagonal), we obtain $M_0$, a non-unitary measurement operator:
\begin{equation}
M_0 = \left(
   \begin{array}{cccc}
   0.849549077650853  &                 0   \\
                   0  & 0.527509587270776
   \end{array}
   \right)
\end{equation}

With the matrices above prepared (in an offline process), we follow the proposed sequence of steps.
We start off by applying the unitary operator $R$ to obtain:
\begin{equation}
\ket{d} = 0.000488281308208\ket{0} - 0.999999880790675\ket{1}
\end{equation}

We go now into the non-unitary phase, where $M_0$ is to be applied 16 times. Consider the effect of just the first iteration.
Note that the success probability in applying $M_0$ is 0.278266470393446.
After applying $M_0$, state becomes:
\begin{equation}
\ket{d} = 0.000786372165026\ket{0} - 0.999999690809361\ket{1}
\end{equation}

We assume for now (in this paper) that $M_0$ is successful throughout the non-unitary phase.
Note that the success probability in applying $M_0$ in this particular example is in the range from about 0.28 to about 0.40.

At the end of the 16 applications of $M_0$, the state becomes:
\begin{equation}
\ket{d} = 0.707106781186547\ket{0} - 0.707106781186548\ket{1}
\end{equation}

Applying $Q$, the state finally becomes:
\begin{equation}
\ket{d} = 0\ket{0}  + 1 \ket{1}
\end{equation}

\section{Notes}

We make a few notes with regard to the method proposed in this paper.
\begin{enumerate}
\item
Quantum search using binary trees has been considered in other works, for example \cite{hoyer_neerbek2001}.
However, the binary tree search proposed in this paper does not assume sorted data.
\item
An alternative approach for differentiating between $\ket{x}$ and $\ket{y}$ is described by Ohya and Volovich in \cite{Ohya2002}. They propose the use of what they call a \textit{chaotic amplifier} to separate the two states.
The method proposed in this paper, however, should be easier to understand and implement.
\item
While it has been proven before that a black-box quantum query algorithm cannot solve \textit{NP} problems in $o(2^{n/2})$ \cite{bennett1997}, 
the approach in this paper incorporates non-unitary model, and hence it does not violate the result in \cite{bennett1997}.
\item
The main catch in the proposed method is that in the state differentiation module, we have not considered the case when $M_1$ instead of $M_0$ was applied during the non-unitary phase, resulting in the intermediate result $\ket{d'}$ rather than $\ket{d}$.
In some cases, $\ket{d'}$ is far enough from $\ket{d}$ so as to be easily distinguishable.
In general, some sort of statistical scheme would be needed to ensure that the final result used is one obtained without or with minimal application of $M_1$.
\end{enumerate}

\bibliography{quantum}

\begin{thebibliography}{10}

\bibitem{bennett1997}
C.H. Bennett, E.~Bernstein, G.~Brassard, and U.~Vazirani.
\newblock Strengths and weaknesses of quantum computing.
\newblock arXiv:quant-ph/9701001.

\bibitem{brassard98}
G.~Brassard, P~H{\o}yer, and A.~Tapp.
\newblock Quantum counting.
\newblock In {\em Proceedings of 25th International Colloquium on Automata,
  Languages and Programming, Lecture Notes in Computer Science}, volume 1443,
  pages 820--831. Springer-Verlag, July 1998.

\bibitem{Cerf1998}
N.J. Cerf, L.K. Grover, and C.P. Williams.
\newblock Nested quantum search and np-complete problems, June 1998.
\newblock arXiv:quant-ph/9806078v1.

\bibitem{durr96}
C.~Durr and P.~H{\o}yer.
\newblock A quantum algorithm for finding the minimum, July 1996.
\newblock arXiv:quant-ph/9607014v2.

\bibitem{grover1996}
L.~Grover.
\newblock A fast quantum mechanical algorithm for database searching.
\newblock In {\em ACM Symposium on Theory of Computing}, pages 212--219. ACM,
  1996.

\bibitem{grover97}
L.~Grover.
\newblock Quantum computers can search rapidly by using almost any
  transformation.
\newblock In {\em Physical Review Letters}, volume~80, pages 4329--4332, May
  1998.

\bibitem{grover05}
L.~Grover.
\newblock A different kind of quantum search, March 2005.
\newblock arXiv:quant-ph/0503205v1.

\bibitem{grover08}
L.~K. Grover.
\newblock Superlinear amplitude amplification, June 2008.
\newblock arXiv:0806.0154v1.

\bibitem{hoyer_neerbek2001}
P.~H{\o}yer, J.~Neerbek, and Y.~Shi.
\newblock Quantum complexities of ordered searching, sorting and element
  distinctness, April 2001.
\newblock arXiv:quant-ph/0102078v2.

\bibitem{nayak1998}
A.~Nayak and F.~Wu.
\newblock The quantum query complexity of approximating the median and related
  statistics, November 1998.
\newblock arXiv:quant-ph/9804066v2.

\bibitem{Ohya2002}
M.~Ohya and I.V. Volovich.
\newblock A new quantum algorithm for np-complete problems, September 2002.
\newblock CDMTCS Research Report Series.

\bibitem{terashima2005}
H.~Terashima and M.~Ueda.
\newblock Nonunitary quantum circuit, April 2005.
\newblock arXiv:quant-ph/0304061v5.

\bibitem{tucci2010}
R.R. Tucci.
\newblock An adaptive, fixed-point version of grover's algorithm, September
  2010.
\newblock arXiv:1001.5200v2.

\end{thebibliography}
\end{document}